\documentclass[12pt,showpacs,preprintnumbers,tightenlines,superscriptaddress,amsmath,amssymb,nofootinbib]{revtex4-1}

\usepackage{graphicx}
\usepackage{amsmath}
\usepackage{amssymb}
\usepackage{bm}
\usepackage{multirow}
\usepackage[caption=false]{subfig}
\usepackage{feynmp}
\usepackage{scalerel}
\usepackage{stackengine}
\stackMath
\usepackage[utf8]{inputenc}
\allowdisplaybreaks[1]

\renewcommand{\ol}{\overline}
\newcommand{\wt}{\widetilde}
\newcommand{\bea}{\begin{eqnarray}}
\newcommand{\eea}{\end{eqnarray}}

\newcommand\fverbdo{\egroup\medskip\noindent%
			\fbox{\unhbox\fverbbox}\ }
\newcommand\fverbit{\egroup\item[\fbox{\unhbox\fverbbox}]}
\newbox\fverbbox

\newcommand{\al}{\alpha}
\newcommand{\be}{\beta}
\newcommand{\ga}{\gamma}
\newcommand{\la}{\lambda}

\newcommand{\De}{\Delta}
\newcommand{\Ga}{\Gamma}

\newcommand{\GeV}{\ \mathrm{GeV}}
\newcommand{\TeV}{\ \mathrm{TeV}}

\newcommand{\scalar}{\phi}
\newcommand{\pseudoscalar}{\widetilde{\phi}}
\newcommand{\maybepseudoscalar}{%
  \stackon[-0.58ex]{\pseudoscalar}
  {\hspace*{0.38ex}\scaleto{\scriptscriptstyle(\hspace*{1.5ex})}{0.85ex}}}
\newcommand{\subscriptmaybepseudoscalar}{%
  \hspace*{-0.3ex}%
  \raisebox{-0.41ex}[0ex]{\scalebox{0.69}{$\maybepseudoscalar$}}}
\newcommand{\cgamma}{c_\gamma}
\newcommand{\tildecgamma}{\widetilde{c}_\gamma}
\newcommand{\maybetildecgamma}{%
  \stackon[-0.58ex]{\tildecgamma}
  {\hspace*{-.75ex}\scaleto{\scriptscriptstyle(\hspace*{1.5ex})}{0.85ex}}}
\newcommand{\maybetildecgammaLHC}{%
  \maybetildecgamma^%
  {\raisebox{-0.38ex}[0pt]{\scriptsize\rm\hspace*{-1.5ex} LHC}}}

\newcommand{\tildefieldstr}[2]{\widetilde{#1}#2}
\newcommand{\maybetildefieldstr}[2]{%
  \stackon[-0.58ex]{\tildefieldstr{#1}{#2}}
  {\hspace*{-1.58ex}\scaleto{\scriptscriptstyle(\hspace*{1.5ex})}{0.85ex}}}
\newcommand{\amu}{a_\mu}
\newcommand{\tildeamu}{\widetilde{a}_\mu}
\newcommand{\maybetildeamu}{%
  \stackon[-0.58ex]{\tildeamu}
  {\hspace*{-1.1ex}\scaleto{\scriptscriptstyle(\hspace*{1.5ex})}{0.85ex}}}

\begin{document} 
\allowdisplaybreaks[2]

\title{LHC 750 GeV diphoton excess and muon \boldmath$(g-2)$}

\preprint{KIAS--Q16003}

\author{Seungwon Baek}
\email{swbaek@kias.re.kr}
\affiliation{School of Physics, KIAS, 85 Hoegiro, Seoul 02455, Korea}

\author{Jae-hyeon Park}
\email{jhpark@kias.re.kr}
\affiliation{Quantum Universe Center, KIAS, 85 Hoegiro, Seoul 02455, Korea}



\begin{abstract}
We consider implications of the diphoton excess recently observed at the LHC on the anomalous magnetic dipole
moment of the muon $(g-2)_\mu = 2 a_\mu$, hypothesizing that the possible 750 GeV resonance is a (pseudo)scalar particle
$\maybepseudoscalar$.
The $\maybepseudoscalar$-$\gamma$-$\gamma$
interaction implied by the diphoton events might generically contribute to $a_\mu$
via 2-loop Barr-Zee type diagrams in a broad class of models.
If $\maybepseudoscalar$ is an $SU(2)_L$ singlet,
the new contribution to $a_\mu$ is much smaller than
the current anomaly,
$\Delta a_\mu\equiv a_\mu^{\rm exp}-a_\mu^{\rm SM} \approx (30 \pm 10) \times 10^{-10}$,
since the scalar can complete the Barr-Zee diagrams only through its mixing with the
Standard Model Higgs boson.
If $\maybepseudoscalar$ belongs to an $SU(2)_L$ doublet in an extended Higgs sector,
then by contrast,
$\Delta a_\mu$ can be easily accommodated with the aid of
an enhanced Yukawa coupling of $\maybepseudoscalar$ to the muon such as in the Type-II or -X two Higgs doublet model.
\end{abstract}
\maketitle

Recently, both the ATLAS and the CMS collaborations at the LHC observed a possible resonance around 750
GeV in the diphoton mass distribution from the dataset of $pp$ collisions at $\sqrt{s}=13 \TeV$ \cite{ATLAS:2015,CMS:2015}.
The significance of deviation reported by ATLAS (CMS) is 3.9 (2.6)~$\sigma$ out of the 3.2 (2.6) fb$^{-1}$ sample if 
the look-elsewhere effect is not taken into account.
The measured excess in the cross section is
\bea
 \sigma(pp \to \gamma\gamma) = \left \{
\begin{array}{ll}
(10 \pm 3) \,{\rm fb} & \quad\text{ATLAS} \\
(6 \pm 3) \,{\rm fb} & \quad\text{CMS}. 
\end{array}
\right.
\label{eq:sigma pp to gaga}
\eea
Clearly, more data are called for
to confirm or exclude this intriguing hint at new physics.

In the meantime, a large number of works have already appeared to put forth
diverse theories on 
a yet-unknown resonance \cite{
DiChiara:2015vdm,Altmannshofer:2015xfo,Djouadi:2016eyy,
singlet,
2HDM VLF,2HDM,Han:2016bvl,Badziak:2015zez,
Franceschini:2015kwy,
Goertz:2015nkp,
Bauer:2015boy,
Nomura:2016fzs,
others diphoton}.
%
From this sharp burst of endeavours,
an outstanding property has emerged
which is common in the majority of the phenomenological models:
the resonance candidate is required to have a rather strong interaction with
a pair of photons to reproduce the experimentally preferred event rate
\cite{Franceschini:2015kwy}.
For instance, the same type of triangle diagrams as
for the Standard Model (SM) Higgs decay into two photons would not suffice.
This has naturally led us to think of a possible connection between
the diphoton excess and another popular observable susceptible to
new electromagnetic interactions, i.e.\
the anomalous magnetic moment of the muon $(g-2)_\mu = 2a_\mu$.
It is well known to show a long-standing deviation of about 3$\sigma$ from the SM prediction~\cite{g-2}:
\bea
\Delta a_\mu \equiv a_\mu^{\rm exp} - a_\mu^{\rm SM} = (29.0 \pm 9.0 \quad \text{to} \quad 33.5 \pm 8.2) \times 10^{-10}.
\label{eq:Del_amu}
\eea
Our aim shall then be to see whether or not this discrepancy can be ameliorated
by generic properties of the newly introduced resonance.


To this end, we shall focus on a broad class of models in which
the 750 GeV resonance is a spin-zero boson.
The pivotal point here is that the (pseudo)scalar-diphoton vertex necessary to produce the excess could generically
contribute to the Barr-Zee type diagrams \cite{Barr:1990vd}
for $a_\mu$~\cite{Arhrib:2001,Baek:2002,Baek:2004tm,Baek:2014,Ilisie:2015,Abe:2015oca},
provided that the resonance couples to the muon.
For this Yukawa coupling,
two mechanisms are conceivable: indirect or direct.
The former is presumably most generic in the sense that it would allow
the boson to interact with muons even if
the muon has a Yukawa coupling only with the SM Higgs doublet.
Prime examples in which this is the case would be
singlet-extensions of the SM
(see e.g.~\cite{%
DiChiara:2015vdm,Altmannshofer:2015xfo,Djouadi:2016eyy,
singlet}).
In this class of models,
the gauge symmetry forbids a renormalizable coupling between
a new $SU(2)_L$-singlet scalar and muons.
It is well known nevertheless that a heavy mass eigenstate can
couple to muons through the mixing between the singlet and the SM Higgs boson.
This leads however to
the drawback that the coupling is suppressed by the mixing angle.
The latter, direct mechanism is more straightforward.
One may simply introduce an additional $SU(2)_L$-doublet which can form
a Yukawa coupling with a muon pair.
Obvious examples include two Higgs doublet models (2HDMs)
(see e.g.~\cite{%
DiChiara:2015vdm,Altmannshofer:2015xfo,Djouadi:2016eyy,
2HDM VLF,2HDM,Han:2016bvl,Badziak:2015zez}).
An important feature of this class of models is that
the resonance-muon coupling can be stronger than
the SM muon Yukawa coupling
depending on the structure of the Higgs-Yukawa sector,
of which one might take advantage to explain $\De a_\mu$.
In 2HDMs for instance, this would amount to
playing with the Higgs mixing angles and the Yukawa ``types''.

In what follows, we evaluate the Barr-Zee type $a_\mu$ diagrams
induced by the resonance-diphoton effective vertex embedded in them.
We consider elementary cases where
the effective vertex is dominated by
a single particle circulating in the loop,
which allows us to obtain a simple
relation between $\De a_\mu$ and the diphoton decay amplitude.
By using the resonance-diphoton coupling strength
preferred by the ATLAS and CMS data,
we predict the range of $\De a_\mu$.
Furthermore,
we comment on a popular case where
the effective vertex arises from vector-like fermions
which can form multiple states in the loop.

Shortly after the announcement from the LHC,
a paper appeared which included qualitative discussion of
the Barr-Zee type contributions to $a_\mu$ mediated by a 750 GeV resonance
\cite{Goertz:2015nkp}.
In our work, we perform a more quantitative analysis
based on concrete prescriptions.
Another paper included the Barr-Zee graphs
although they did not play a significant role in the results
\cite{Han:2016bvl}.
Apart from the Barr-Zee diagrams,
other types of corrections to $a_\mu$ might arise
which depend on the details of each model for the diphoton excess
\cite{Bauer:2015boy,Djouadi:2016eyy,Nomura:2016fzs}.



We begin the analysis
by introducing an effective Lagrangian~\cite{Falkowski:2012} to describe
the decay of a 750 GeV (pseudo)scalar $\maybepseudoscalar$ into two photons:
\bea
{\cal L}_{\rm eff} =
  \maybetildecgamma\frac{\alpha}{\pi v}
  \maybepseudoscalar F_{\mu\nu} \maybetildefieldstr{F}{^{\mu\nu}}
,
\label{eq:Leff}
\eea
where the electromagnetic dual field strength tensor is given by
$\tildefieldstr{F}{^{\mu\nu}} = \epsilon^{\mu\nu\rho\sigma} F_{\rho\sigma} / 2$
and $v\approx 246$ GeV is the vacuum expectation value (VEV)
of the SM Higgs field.
One can express the effective coupling strength $|\maybetildecgamma|$
needed to fit the LHC diphoton excess in the form
\bea
|\maybetildecgammaLHC| \simeq 5.0\times \left(\Ga_{\ga\ga} / m_{\subscriptmaybepseudoscalar} \over 1.0 \times 10^{-4} \right)^{1/2} ,
\label{eq:cgamma}
\eea
in terms of the decay width
$\Gamma_{\gamma\gamma} \equiv \Gamma(\maybepseudoscalar \to\gamma\gamma)$.
This relation allows us to
estimate a viable range of $|\maybetildecgamma|$ from that of
$\Gamma_{\gamma\gamma}$,
\bea
1.1 \times 10^{-6} \lesssim \Gamma_{\gamma\gamma}/m_{\subscriptmaybepseudoscalar} \lesssim 2 \times 10^{-3},
\label{eq:range Gamma gamma gamma}
\eea
determined from its dependence on
$\Gamma_{gg} \equiv \Gamma(\maybepseudoscalar \to g g)$
as well as $\sigma(pp \to \maybepseudoscalar \to \gamma\gamma)$
\cite{Franceschini:2015kwy}.
The lower limit arises from the condition that $\maybepseudoscalar\to\gamma\gamma$ and $\maybepseudoscalar\to g g$ saturate the total width $\Gamma_{\rm tot}$
while the upper limit applies when
the $\maybepseudoscalar$ production is dominated by photon fusion.

Values of $|\maybetildecgammaLHC|$ from~\eqref{eq:cgamma}
and~\eqref{eq:range Gamma gamma gamma} are much larger than the size
of $\cgamma$ which would result if
the 750 GeV scalar had only SM-like interactions.
More concretely, this coupling would read
\begin{equation}
  \cgamma^\mathrm{SML} = \cgamma^\mathrm{SML}(f) + \cgamma^\mathrm{SML}(V),
  \label{eq:cgammaSML}
\end{equation}
where the fermion and the vector-boson loop contributions,
\begin{subequations}
  \label{eq:cgamma top W}
  \begin{align}
\cgamma^\mathrm{SML}(f) &=
{N(r_t) Q_t^2  \over 6} A_f(\tau_t) , \\
\cgamma^\mathrm{SML}(V) &=
 - {7 \over 8 } A_v(\tau_W) ,
  \end{align}
\end{subequations}
would result from the SM-like top- and $W$-loops.
Here,
$N(r_t) = 3$ is the number of top quarks with different colours,
$Q_t = 2/3$ is the top quark charge,
the loop functions $A_f(\tau)$ and $A_v(\tau)$
are given in~\cite{Falkowski:2012}, and
$\tau_t = m_\phi^2 / 4 m_t^2$,
$\tau_W = m_\phi^2 / 4 m_W^2$.
The numerical size of \eqref{eq:cgammaSML} would then be\footnote{We consider only the scalar case.
For the pseudoscalar case, the vector boson contribution should be omitted.}
\bea
|\cgamma^\mathrm{SML}| \simeq 0.087,
\label{eq:SM-like}
\eea 
which is two orders of magnitude smaller than the typical size of
$\maybetildecgammaLHC$ from~\eqref{eq:cgamma}.

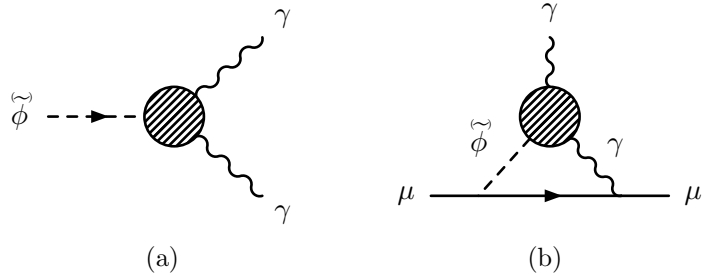
\begin{figure}
  \captionsetup[subfigure]{captionskip=1.5em}
  \subfloat[]{%
    \begin{fmffile}{phi-gamma-gamma}
      \begin{fmfgraph*}(90,60)
        \fmfset{arrow_ang}{20}
        \fmfset{arrow_len}{2.5mm}
        \fmfleft{phi}
        \fmfright{ga1,ga2}
        \fmf{dashes_arrow,tension=1.4}{phi,v1}
        \fmf{photon}{v1,ga1}
        \fmf{photon}{v1,ga2}
        \fmfblob{0.25w}{v1}
        \fmfv{l=$\maybepseudoscalar$,l.a=160}{phi}
        \fmflabel{$\gamma$}{ga1}
        \fmflabel{$\gamma$}{ga2}
      \end{fmfgraph*}
    \end{fmffile}%
  }
  \qquad\qquad
  \subfloat[]{%
    \begin{fmffile}{barr-zee}
      \begin{fmfgraph*}(90,60)
        \fmfset{arrow_ang}{20}
        \fmfset{arrow_len}{2.5mm}
        \fmfstraight
        \fmfbottom{muL,muR}
        \fmftop{ga1}
        \fmf{plain,tension=3}{muL,phi}
        \fmf{plain_arrow}{phi,ga2}
        \fmf{plain,tension=3}{ga2,muR}
        \fmffreeze
        \fmf{dashes,label=$\maybepseudoscalar$}{v1,phi}
        \fmf{photon,label=$\gamma$}{v1,ga2}
        \fmf{photon,tension=2}{v1,ga1}
        \fmfblob{0.25w}{v1}
        \fmfv{l=$\gamma$,l.a=90}{ga1}
        \fmfv{l=$\mu$,l.a=180}{muL}
        \fmfv{l=$\mu$,l.a=0}{muR}
      \end{fmfgraph*}
    \end{fmffile}%
  }
\caption{Generic diagram for $\maybepseudoscalar\to\gamma\gamma$ (a) and Barr-Zee type diagram for $a_\mu$ (b).}
\label{fig:phirr}
\end{figure}

This requires contributions to $\maybetildecgamma$
much bigger than $\cgamma^\mathrm{SML}$,
presumably arising from certain underlying physics.
We sketch a generic diagram for this in Fig.~\ref{fig:phirr}(a).
This graph can be embedded in Fig.~\ref{fig:phirr}(b)
thereby generating a contribution to $a_\mu$,
provided that $\maybepseudoscalar$ couples to muons.

In view of the large $|\maybetildecgamma|$ from~(\ref{eq:cgamma}),
it might be expected to induce a sizeable $\Delta a_\mu$.
An obstacle is however that the effective operator in~(\ref{eq:Leff}) cannot be used for a direct calculation of the diagram in Fig.~\ref{fig:phirr}(b).
This is mainly due to the difference between the kinematics involved in the $\maybepseudoscalar \to \gamma\gamma$ and $\Delta a_\mu$ calculations.
For example, the photons in Fig.~\ref{fig:phirr}(a) are highly energetic while the external photon in
Fig.~\ref{fig:phirr}(b) is very soft.  A naive application of (\ref{eq:Leff}) would lead to both ultraviolet and infrared
divergences in $\Delta a_\mu$.

To circumvent these problems,
we shall assume that the $\maybepseudoscalar$-$\gamma$-$\gamma$ vertex
originates from loops of heavy particles.
Moreover, we shall mainly focus on cases where the effective vertex is
dominated by a one-loop contribution involving a single particle,
to make the $\Delta a_\mu$ predictions as model-independent as possible.
As we will see,
relaxing this single-particle dominance would lead to similar conclusions.
We shall consider three types of particles that could appear inside the loop in Fig.~\ref{fig:phirr}(a): fermion ($f$), vector ($V$), and scalar ($S$).
Their interaction Lagrangian might read
\bea
{\cal L} =
-\xi_\phi^f {m_f \over v} \phi \ol{f} f
+ \rho^V_\phi {2 m_V^2 \over v} \phi V_\mu^\dagger V^\mu
-\la_{\phi}^S v \phi S^\dagger S
+i \xi_{\pseudoscalar}^f {m_f \over v} \pseudoscalar\,\ol{f} \ga_5 f.
\label{eq:Lint}
\eea
The individual contributions to $\maybetildecgamma$ can then be written in the forms~\cite{Falkowski:2012},
\begin{subequations}
  \label{eq:cgamma phi}
  \begin{align}
\cgamma(f) &= 
{N(r_f) Q_f^2 \xi_\phi^f \over 6} A_f(\tau_f) , \\
\cgamma(V) &= 
-{7 N(r_V) Q_V^2 \rho_\phi^V \over 8} A_v(\tau_V) , \\
\cgamma(S) &= 
{N(r_S) Q_S^2 \la_\phi^S v^2 \over 48 m_S^2} A_s(\tau_S) , \\
\tildecgamma(f) &=
-{N(r_f) Q_f^2 \xi_{\pseudoscalar}^f \over 4} A_a(\tau_f) ,
  \end{align}
\end{subequations}
where $\tau_i = m_{\subscriptmaybepseudoscalar}^2/4 m_i^2$ with $i=f,V,S$, and the loop functions $A_{f,v,s}(\tau)$ and $A_a(\tau)$ are given in~\cite{Falkowski:2012,Djouadi:1993ji}, respectively.
For each of these four types of vertices,
one can evaluate the corresponding two-loop Barr-Zee graph
in Fig.~\ref{fig:phirr}(b) to obtain the generic formulae~\cite{Arhrib:2001,Ilisie:2015},
\begin{subequations}
  \begin{align}
\De a_\mu(f) &= \frac{\al m_\mu^2}{4 \pi^3 v^2} N(r_f) Q_f^2 \xi_\phi^f \xi_\phi^\mu {\cal F}_f (z_{f\phi}), \\
\De a_\mu(V) &= \frac{\al m_\mu^2}{8 \pi^3 v^2} N(r_V) Q_V^2 \rho_\phi^V \xi_\phi^\mu {\cal F}_v (z_{V\phi}), \\
\De a_\mu(S) &= \frac{\al m_\mu^2}{8 \pi^3 m_S^2} N(r_S) Q_S^2 \la_\phi^S \xi_\phi^\mu {\cal F}_s (z_{S\phi}),
\\
\De \wt{a}_\mu(f) &= \frac{\al m_\mu^2}{4 \pi^3 v^2} N(r_f) Q_f^2 \xi_{\pseudoscalar}^f \xi_{\pseudoscalar}^\mu \wt {\cal F}_f (z_{f\pseudoscalar}), 
  \end{align}
\end{subequations}
where $z_{i\subscriptmaybepseudoscalar}=m_i^2/m_{\subscriptmaybepseudoscalar}^2$ with $i=f,V,S$, and the loop functions are given by
\begin{subequations}
  \begin{align}
{\cal F}_f(z) &= {z \over 2} \int_0^1 dx \frac{2x(1-x)-1}{z-x(1-x)} \log\frac{z}{x(1-x)}, \\
{\cal F}_v(z) &= {1 \over 2} \int_0^1 dx \frac{x(12x^2-3x+10)z-x(1-x)}{z-x(1-x)} \log\frac{z}{x(1-x)}, \\
{\cal F}_s(z) &= {z \over 2} \int_0^1 dx \frac{x(x-1)}{z-x(1-x)} \log\frac{z}{x(1-x)}, \\
\wt{\cal F}_f(z) &= {z \over 2} \int_0^1 dx \frac{1}{z-x(1-x)} \log\frac{z}{x(1-x)}.
  \end{align}
\end{subequations}

With the above ingredients,
we are ready to consider the first class of models, i.e.\
those with an additional scalar field $\Phi$ (complex or real)
which is singlet under the SM gauge group.
Even though the gauge symmetry forbids a direct $\Phi$-$\mu$-$\mu$ coupling,
a renormalizable interaction of the form,
\bea
    \lambda_{H\Phi} H^\dagger H \Phi^\dagger \Phi,
\eea
can mix $\phi$, a real degree of freedom out of $\Phi$,
with $h$, the SM Higgs from the Higgs doublet $H$,
if both $H$ and $\Phi$ acquire VEVs.
Barring $CP$-violation,
$\phi$ here should be a $CP$-even scalar.
The lighter and the heavier mass eigenstates $H_1$ and $H_2$ can then be
identified with the 125 GeV Higgs boson and the 750 GeV resonance, respectively.
We let $R(\theta)$ denote the mixing matrix
\begin{equation}
\setlength\arraycolsep{0.7ex}
R(\theta) =
\left(
\begin{array}{rr}
 \cos\theta & -\sin\theta \\
 \sin\theta &  \cos\theta
\end{array}
\right) ,
\label{eq:R}
\end{equation}
so that
\bea
\left(
\begin{array}{c}
h \\
\phi
\end{array}
\right)
= R(-\alpha)
\left(
\begin{array}{c}
H_1 \\
H_2
\end{array}
\right)
.
\label{eq:mixing}
\eea
In terms of the mixing angle $\alpha$,
the three pieces of couplings for $H_2\to \gamma\gamma$ are written as
\begin{subequations}
  \label{eq:cgammaH2}
  \begin{align}
\cgamma^{H_2}(f)&= c_\al \cgamma(f)
 + s_\al \cgamma^\mathrm{SML}(f), \\
\cgamma^{H_2}(V)&= c_\al \cgamma(V)
 + s_\al \cgamma^\mathrm{SML}(V), \\
\cgamma^{H_2}(S)&= c_\al \cgamma(S),
  \end{align}
\end{subequations}
where the $\phi$-components and the SM-like components are
from~\eqref{eq:cgamma phi} and~\eqref{eq:cgamma top W}, respectively.

\begin{figure}[tbh]
\includegraphics[width=0.6\textwidth]{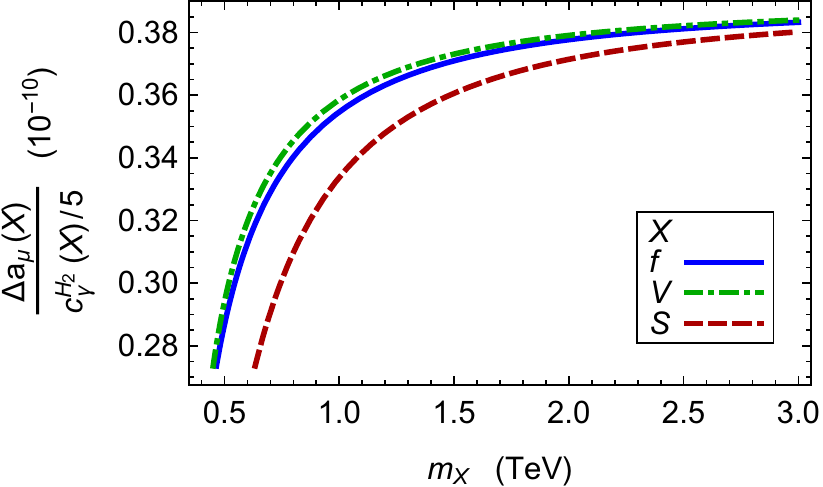}
\caption{The ratio $\De a_\mu(X)/\cgamma^{H_2}(X)$
  as a function of $m_X$, mass of the particle $X$
  in the $\scalar$-$\gamma$-$\gamma$ loop,
  for $X = f,V,S$, standing for
  fermion (solid), vector (dash-dotted), or scalar (dashed),
  respectively.}
\label{fig:ratio}
\end{figure}

Since the SM result of the Higgs decay into two photons agrees with the LHC data \cite{LHCHiggs}, $R$ should be close
to a unit matrix with $|\alpha| \ll 1$.  Given that the SM-like contribution $\cgamma^\mathrm{SML}$ has a small size $\sim 0.087$ [see
(\ref{eq:SM-like})],
we can safely neglect the terms suppressed by $s_\al$ in (\ref{eq:cgammaH2}).
For each particle type of fermion, vector, and scalar,
dominating the $\phi$-$\gamma$-$\gamma$ loop,
we then get a ratio of $\De a_\mu$ to $\cgamma^{H_2}$
in one of the simple forms,
\begin{subequations}
  \begin{align}
\frac{\De a_\mu(f)}{\cgamma^{H_2}(f)} &\simeq -\frac{3 \al s_\al m_\mu^2}{2 \pi^3 v^2} 
\frac{{\cal F}_f(z_{fH_1}) -{\cal  F}_f(z_{fH_2})}{A_f(\tau_f)}, \\
\frac{\De a_\mu(V)}{\cgamma^{H_2}(V)} &\simeq \frac{ \al s_\al m_\mu^2}{7 \pi^3 v^2} 
\frac{{\cal F}_v(z_{VH_1}) -{\cal F}_v(z_{VH_2})}{A_v(\tau_V)}, \\
\frac{\De a_\mu(S)}{\cgamma^{H_2}(S)} &\simeq -\frac{6 \al s_\al m_\mu^2}{ \pi^3 v^2} 
\frac{{\cal F}_s(z_{SH_1}) -{\cal F}_s(z_{SH_2})}{A_s(\tau_S)}. 
  \end{align}
\end{subequations}
We can see that many parameters have been cancelled out of the numerator and denominator.
In Fig.~\ref{fig:ratio}, we plot the above ratios as functions of $m_X$ ($X=f,V,S$), the particle mass in the $\phi$-$\gamma$-$\gamma$ loop.
We fixed $s_\al=0.1$, a representative small mixing
which is still allowed \cite{Chpoi:2013wga}.
It is interesting to
find that the ratios are similar to one another especially in
the decoupling regime. Given that $|\maybetildecgammaLHC|$ in (\ref{eq:cgamma}) is around 5 we can \emph{predict}
$\De a_\mu$ to be around $\text{a few}\times 10^{-11}$ or less. This prediction is however too small to explain the deviation
in (\ref{eq:Del_amu}).

\begin{figure}[tbh]
\includegraphics[width=0.6\textwidth]{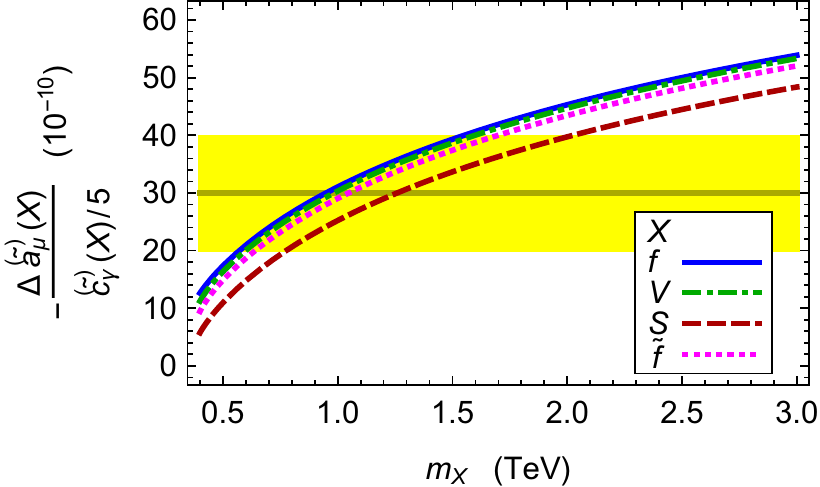}
\caption{The ratio $\De a_\mu(X)/\cgamma(X)$
  as a function of $m_X$, mass of the particle $X$
  in the $\scalar$-$\gamma$-$\gamma$ loop,
  for $X = f,V,S$, standing for
  fermion (solid), vector (dash-dotted), or scalar (dashed),
  respectively, as well as
  $\De \tildeamu(f)/\tildecgamma(f)$ as a function of $m_f$ (dotted).
  The horizontal band depicts the experimentally preferred range of
  $\De a_\mu$.}
\label{fig:ratioH}
\end{figure}

Now we turn to the second class of models, in which the Higgs sector includes new scalars transforming non-trivially under the $SU(2)_L$ group.
Let us consider a 2HDM as the simplest example. Our results could be easily extended to more complicated models with
additional doublets. In 2HDMs, a $Z_2$ symmetry is often imposed to prevent dangerous flavour changing neutral
currents mediated by Higgs at tree-level \cite{Weinberg:1976hu}. Depending on the assignment of the $Z_2$ parity, there are four different types of naturally flavour-conserving (NFC) models \cite{Glashow:1976nt},
named, Type-I, -II, -X, and -Y (see e.g.~\cite{Aoki:2009ha}).
In the Type-II and Type-X models,
the Yukawa coupling of the heavier (pseudo)scalar $H$($A$) to leptons can be enhanced by the factor $\xi_{H(A)}$.
Both these factors become $\tan\be$ in
the alignment limit with $\sin(\be-\al)=1$, which we shall adopt
for the lightest state to have SM-like properties \cite{Craig:2012vn}.
This also makes it simpler to consider heavy Higgs decays by
suppressing the $H$ couplings to vector bosons and to $hh$.
If $H$ or $A$ is identified with the 750 GeV resonance, the ratio $\De \maybetildeamu/\maybetildecgamma$ can be written approximately as
\begin{subequations}
  \begin{align}
\frac{\De a_\mu(f)}{c_\ga(f)} &\simeq \frac{3 \al \xi_H^\mu m_\mu^2}{2 \pi^3 v^2} \frac{{\cal F}_f(z_{fH})}{A_f(\tau_f)}, \\
\frac{\De a_\mu(V)}{c_\ga(V)} &\simeq -\frac{\al \xi_H^\mu m_\mu^2}{7 \pi^3 v^2} \frac{{\cal F}_v(z_{VH})}{A_v(\tau_V)}, \\
\frac{\De a_\mu(S)}{c_\ga(S)} &\simeq \frac{6 \al \xi_H^\mu m_\mu^2}{ \pi^3 v^2} \frac{{\cal F}_s(z_{SH})}{A_s(\tau_S)}, \\
\frac{\De \wt{a}_\mu(f)}{\wt{c}_\ga(f)} &\simeq -\frac{\al \xi_A^\mu m_\mu^2}{\pi^3 v^2} \frac{\wt{\cal F}_f(z_{fA})}{A_a(\tau_f)},
  \label{eq:ratio fA}
  \end{align}
\end{subequations}
keeping only the $\tan\be$-enhanced terms.
Fig.~\ref{fig:ratioH} shows the above ratios as functions of $m_X$,
the particle mass in the $\maybepseudoscalar$-$\gamma$-$\gamma$ loop.
We took $\xi_H^\mu =\xi_A^\mu =\tan\be =10$. It is remarkable that the $\De a_\mu$ in (\ref{eq:Del_amu}) can be easily
explained with $|\maybetildecgammaLHC| \sim 5$ in (\ref{eq:cgamma}) if $\maybetildecgamma <0$.  For this, one could make the sign of each piece of $\maybetildecgamma$ negative by choosing appropriate signs of the coupling constants appearing in~\eqref{eq:Lint}. We also note that the ratios have the same sign and
similar magnitudes. This implies enough possibility that the ratio of total $\De a_\mu$ to total $\maybetildecgamma$ remains somewhere among the curves in Fig.~\ref{fig:ratioH}
even in the presence of several simultaneous contributions from $f$, $V$, and $S$, with comparable sizes.

Since we rely on the enhancement of
the $\maybepseudoscalar$-$\mu$-$\mu$ coupling, remarks on the decays of
$\maybepseudoscalar$ into lepton pairs are in order
\cite{Badziak:2015zez,Abe:2015oca}.
The expression for
$\Gamma_{\ell\ell} \equiv \Gamma(\maybepseudoscalar \rightarrow \ell^+\ell^-)$
for a lepton $\ell$,
\begin{equation}
\frac{\Gamma_{\ell\ell}}{\Gamma_{\gamma\gamma}} \simeq
\frac{\pi^2 m_\ell^2 \xi_{\subscriptmaybepseudoscalar}^{\ell 2}}
{2 \alpha^2 m_{\subscriptmaybepseudoscalar}^2 \maybetildecgamma^2} ,
\end{equation}
can be put in the semi-numerical form,
\begin{equation}
  \begin{bmatrix}
    {\Gamma_{\mu\mu}}/{\Gamma_{\gamma\gamma}} \\
    {\Gamma_{\tau\tau}}/{\Gamma_{\gamma\gamma}}
  \end{bmatrix}
  \simeq
  \left[
  \begin{array}{l}
7.4\times 10^{-3} \\      
2.1
  \end{array}
  \right]
\biggl( \frac{5}{\maybetildecgamma} \biggr)^2
\biggl( \frac{\xi_{\subscriptmaybepseudoscalar}^{\mu,\tau}}{10}\biggr)^2 .
\label{eq:Gamma ll over Gamma gaga numeric}
\end{equation}
From this, it is obvious that
the bound
$\Gamma_{\mu\mu} / \Gamma_{\gamma\gamma} \lesssim 0.6$ from~\cite{Franceschini:2015kwy}
is fulfilled by the parameters chosen above.
Even if one further assumes that
the $\maybepseudoscalar$-$\tau$-$\tau$ coupling is enhanced by the same factor
as in NFC 2HDMs,
the bound
$\Gamma_{\tau\tau} / \Gamma_{\gamma\gamma} \lesssim 6$ from~\cite{Franceschini:2015kwy}
is still satisfied albeit with a less margin.
One can then multiply \eqref{eq:Gamma ll over Gamma gaga numeric} by
\eqref{eq:sigma pp to gaga} to estimate the dilepton production
cross sections through the resonance at 13 TeV,
\begin{equation}
  \begin{bmatrix}
  \sigma(pp \to \mu^+\mu^-) \\
  \sigma(pp \to \tau^+\tau^-)
  \end{bmatrix}
  \simeq
  \left[
  \begin{array}{r}
0.059 \ \mathrm{fb} \\
17   \ \mathrm{fb}
  \end{array}
  \right]
\biggl( \frac{5}{\maybetildecgamma} \biggr)^2
\biggl( \frac{\xi_{\subscriptmaybepseudoscalar}^{\mu,\tau}}{10}\biggr)^2 ,
\label{eq:sigma pp to mumu tautau}
\end{equation}
where we have taken an average of the two diphoton cross sections.
These channels might be handles to cross check the present proposal
at future runs of the LHC\@.
Notice that
the 13~TeV tau pair cross section might already exceed
12~fb, the 95\% confidence level upper limit from the 8~TeV run
\cite{Aad:2014vgg}.
If heavy Higgs searches in this channel rule out
the tau pair production rate suggested by $\De \amu$
in the near future,
then one might resort to a non-NFC scenario where
$\xi_{\subscriptmaybepseudoscalar}^\mu \neq
\xi_{\subscriptmaybepseudoscalar}^\tau$.

Depending on the model,
the decay $\maybepseudoscalar \to b\bar{b}$
might also be enhanced as is the case e.g.\ in the Type-II 2HDM\@,
in contrast to the Type-X model where the $b\bar{b}$ mode
would be suppressed.
To compensate for the
suppression of $\mathrm{BR}(\maybepseudoscalar \to \gamma\gamma)$,
the former model would therefore require a higher
production rate of $pp \to \maybepseudoscalar$
than the latter.  
We shall leave the construction of
a resonance production mechanism
out of the scope of this work,
as our discussion is independent of
a concrete realization thereof.
In such a model,
it would also be of interest to look for
heavy Higgs decays into $b\bar{b}$,
whose event rate would be roughly 10 times that of the
$\tau^+\tau^-$ channel shown in \eqref{eq:sigma pp to mumu tautau}.

In the Type-II model,
$b \to s \gamma$ places a lower limit on the charged Higgs mass $m_{H^\pm}$
around 480 GeV at the 95\% confidence level \cite{Misiak:2015xwa}.
This is easy to satisfy without spoiling the $\rho$ parameter
by assuming that $m_{H^\pm} \sim m_A \sim 750\GeV$.
To probe an effect from such a heavy charged Higgs
on $b \to s \gamma$, one would need a better experimental precision
than would be available at a super flavour factory
as well as more accurate theory predictions
[see e.g.\ Fig.~7(b) of \cite{Hermann:2012fc}].

The rather large modulus of $\maybetildecgamma$ for a high enough rate of
$\maybepseudoscalar \to \gamma\gamma$ is a challenge to all weakly-coupled
models (see e.g.~\cite{Franceschini:2015kwy}).
This is even more the case if the same type of coupling is to make
a sufficient contribution to $a_\mu$ \cite{Goertz:2015nkp}.
For instance, $\cgamma(f) \sim -5$ was shown above to fit the central value of
$\De a_\mu$.
This would require
$|N(r_f) Q_f^2 \xi_\phi^f| \sim 29$
when $m_f \sim 1\TeV$ for instance,
which may still be considered to lie within the boundary of perturbativity.

A popular way to model massive charged fermions
which couple to heavier Higgses is to make them vector-like
\cite{Badziak:2015zez,Djouadi:2016eyy,Han:2016bvl,Altmannshofer:2015xfo,
  2HDM VLF}.
For instance, a single vector-like ``generation''
might consist of the following fermions:
\begin{equation}
  \begin{array}{llllllll}
    l   & (\mathbf{2},  Q_f + 1/2); &
    l^c & (\mathbf{2},- Q_f - 1/2); &
    e   & (\mathbf{1},  Q_f);       &
    e^c & (\mathbf{1},- Q_f);
  \end{array}
\end{equation}
where each left-handed Weyl spinor is followed by
its $SU(2)_L$ representation and hypercharge enclosed in parentheses.
Suppose that these new fermions have $Z_2$ parities such that
they can couple to one of the Higgs doublets $\Phi_i$
with $i = 1 \text{ or } 2$.
This would allow the following terms in the Lagrangian:
\begin{equation}
  - \mathcal{L} =
    y \Phi_i^\dagger l e^c + y^c \Phi_i^T l^c e
  + m_l l l^c + m_e e e^c
  + \mathrm{h.c.} 
  \label{eq:Lvll}
\end{equation}
With the VEV of $\Phi_i$ taken into account,
the above Weyl fermions comprise
two Dirac mass eigenstates,
each of which can be regarded as $f$ appearing in \eqref{eq:Lint}.
One can obtain the coefficients $\xi_{\subscriptmaybepseudoscalar}^f$ therein
in terms of $y$, $y^c$,
the mass eigenvalues and mixing matrices of the vector-like fermions,
as well as the Higgs mixing angles.
For simplicity,
we shall assume all the parameters to be real-valued in \eqref{eq:Lvll}.
Using mass insertion approximation,
one can then express
the sum of contributions to $\maybetildecgamma$
from the pair of $f$ in the forms,
\begin{subequations}
  \begin{align}
    \cgamma(f) &\simeq
    -
    {N(r_f) Q_f^2 \over 3}
    {R^\beta_{i1} R^\alpha_{i1} v^2 \over m^2_\phi}
    \left[
    (y + y^c)^2 \tau^2 A_f'(\tau) +
    2 y y^c \tau A_f(\tau)
    \right],
    \\
    \setcounter{equation}{3}
    \tildecgamma(f) &\simeq
    \frac{N(r_f) Q_f^2}{2}
    {R^\beta_{i1} R^\beta_{i2} v^2 \over m^2_{\pseudoscalar}}
    (y^{c2} - y^2)\,
    \tau^2 A_a'(\tau) ,
    \label{eq:cgamma VLL pseudo}
  \end{align}
\end{subequations}
where $R^\alpha$ and $R^\beta$ are respectively the abbreviations of
the Higgs mixing matrices
$R(\alpha)$ and $R(\beta)$ as shown in \eqref{eq:R}, and
$\tau = m_{\subscriptmaybepseudoscalar}^2/2 (m_l^2 + m_e^2)$.
A similar operation for $\De \maybetildeamu(f)$ leads to
\begin{subequations}
  \label{eq:amu VLL}
  \begin{align}
    \De a_\mu(f) &\simeq
    \frac{\al m_\mu^2}{4 \pi^3 m_\phi^2}
    N(r_f) Q_f^2 \xi_\phi^\mu R^\beta_{i1} R^\alpha_{i1}
    \left[
    \frac{(y + y^c)^2}{2} \mathcal{F}_f'(z) -
    \frac{y y^c}{z} \mathcal{F}_f(z)
    \right],
    \label{eq:amu VLL scalar}
    \\
    \setcounter{equation}{3}
    \De \wt{a}_\mu(f) & \simeq
    \frac{\al m_\mu^2}{8 \pi^3 m_{\pseudoscalar}^2}
    N(r_f) Q_f^2 \xi_{\pseudoscalar}^\mu R^\beta_{i1} R^\beta_{i2}
    (y^{c2} - y^2) \wt {\cal F}_f' (z),
    \label{eq:amu VLL pseudo}
  \end{align}
\end{subequations}
where
$z = (m_l^2 + m_e^2) / 2m_{\subscriptmaybepseudoscalar}^2$.
The above approximations are valid under the condition that
$m_l \approx m_e \gg \max(|y|, |y^c|) R^\beta_{i1} v$.

Regarding the sign of $\De \tildeamu(f)$,
it is clear from~\eqref{eq:amu VLL pseudo} that
one can make it positive by choosing either $|y|$ or $|y^c|$
to be much larger than the other.
Similarly,
one can check that \eqref{eq:amu VLL scalar} becomes positive
for a pair of $y$ and $y^c$ with comparable magnitudes and appropriate signs.
This is to be contrasted with
vanilla NFC 2HDMs in which the sign of each Barr-Zee contribution
is determined by the type of the Yukawa structure,
the parity of the exchanged scalar,
as well as
whether the fermion in the loop is of up-type or down-type.
For instance, the $H$-$\tau$ contribution to $a_\mu$
in Type-II and -X is negative (see e.g.\ \cite{Chun:2015xfx}).
In a model extended with extra vector-like fermions, as shown above,
there is enough freedom to engineer their gauge quantum numbers
and Yukawa couplings so that
their Barr-Zee contributions, mediated by $H$ or $A$,
have the desired sign.

\begin{figure}
  \subfloat[Scalar exchange, $\tan\be = 10$]{%
    \includegraphics[width=0.6\textwidth]{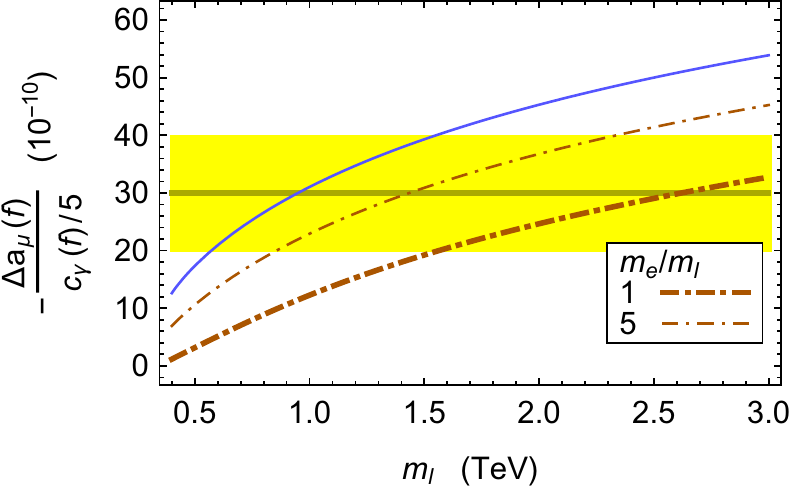}}

  \subfloat[Pseudoscalar exchange, $\tan\be = 1$]{%
    \includegraphics[width=0.6\textwidth]{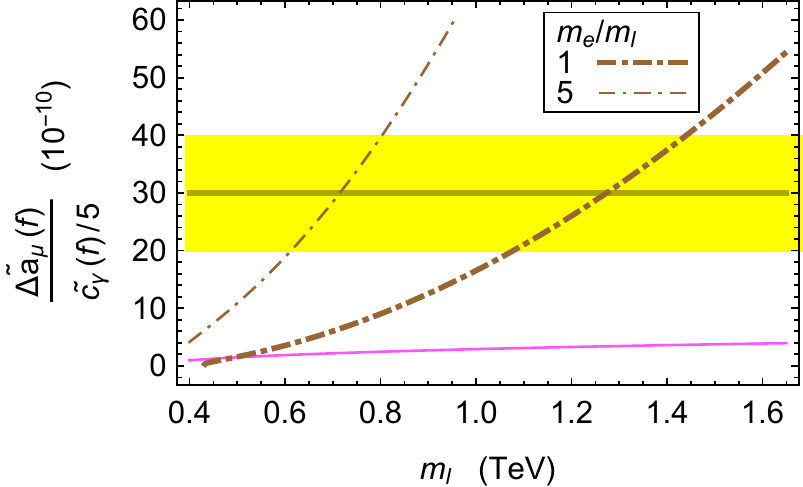}}
  \caption{(a) $\De a_\mu(f)/\cgamma(f)$
    as a function of $m_l$ 
    with $y = y^c$. 
    (b) $\De \tildeamu(f)/\tildecgamma(f)$
    as a function of $m_l$ 
    with
    $y^c = 0$. 
    In both plots, $m_e / m_l$ is fixed to 1 and 5
    on the thick and the thin non-solid curves, respectively.
    The thin solid curves represent the values that would result from
    single-particle dominance.}
  \label{fig:ratios VLL}
\end{figure}

We restrict ourselves hereafter to the $i=1$ case since
coupling $\Phi_2$ to the vector-like fermions would make
an excessive modification to $h \to \gamma\gamma$.
We plot the ratio $\De \maybetildeamu(f)/\maybetildecgamma(f)$ in
Fig.~\ref{fig:ratios VLL} without using mass insertion approximation.
We set each of $y$ and $y^c$ to either 0 or 1
such that $\De \maybetildeamu(f) > 0$.
The overall size of $y^{(c)}$
drops out of the ratio to a good approximation.

In Fig.~\ref{fig:ratios VLL}(a) for scalar exchange,
the curves resemble those in Fig.~\ref{fig:ratioH}
although they lie lower than $\De a_\mu(f)/\cgamma(f)$
in the single-particle case,
reproduced here as the thin solid curve from Fig.~\ref{fig:ratioH}.
This is because
the contributions to each of $\De a_\mu$ and $\cgamma$
from the two mass eigenstates of the vector-like fermions
add up destructively.
One can bring the $i=1$ curve closer to the single-particle result
by decoupling the heavier state, as illustrated by
the thin curve on which $m_e/m_l = 5$.
The plot would remain the same under the interchange of
$m_l \leftrightarrow m_e$ in both the legend and the horizontal axis.

A similar plot for pseudoscalar exchange is shown in
Fig.~\ref{fig:ratios VLL}(b).
Here, the curves have qualitatively different shapes from those in
Fig.~\ref{fig:ratios VLL}.
Another difference is that
the curves indicate much higher values of
$\De \tildeamu(f)/\tildecgamma(f)$
than the single-particle case.
These differences can be traced to the ratio
$\wt {\cal F}_f' (z) / [\tau^2 A_a'(\tau)]$ from
\eqref{eq:amu VLL pseudo} and \eqref{eq:cgamma VLL pseudo},
as opposed to
${\wt{\cal F}_f(z)} / {A_a(\tau)}$
appearing in \eqref{eq:ratio fA}.

On the whole, Fig.~\ref{fig:ratios VLL} reveals a chance
to explain $\De \amu$ as well as the diphoton events
by extending a 2HDM with vector-like fermions.
Even though the curves therein look different from
those in Fig.~\ref{fig:ratioH}, it is still possible to find points
with the best value of the ratio.
Generalization to cases with multiple generations should be
straightforward as long as those generations are
arranged in such a way that their contributions add up constructively.
To fit the preferred size of $\De \amu$,
the above particular model would in practice require
a large number of generations, a large $Q_f$, and/or a large $|y^{(c)}|$,
which might threaten the calculability of the model.
In this respect, the scalar exchange contributions 
seem to be more promising. 
For $m_l = m_e \sim 1\TeV$,
one can account for $\De \amu$ and $\cgamma$ by setting
$Q_f \sim 23$, $y = y^c \sim 4$, and $\tan\be \sim 25$,
which is still around the boundary of the calculability criteria
from (3.25) of \cite{Goertz:2015nkp} with $N_E = 3$.
Note that the pseudoscalar exchange contributions are suppressed
in this case [see \eqref{eq:cgamma VLL pseudo} and \eqref{eq:amu VLL pseudo}].
The high value of $\tan\be$ was chosen to suppress
the extra fermion loop contribution to $h \to \gamma\gamma$
below $2 \sigma$ \cite{LHCHiggs}.
This would however be disfavoured by the limit on $\Gamma_{\tau\tau}$
in a NFC model [see \eqref{eq:Gamma ll over Gamma gaga numeric}].
As mentioned before,
one might consider a non-NFC model in such a case.


To sum up,
we attempted to predict generic contributions to the muon anomalous magnetic moment
from a spin-zero particle $\maybepseudoscalar$ which we assume to produce
the recently observed 750 GeV diphoton events. We evaluated the ratio of
each component of $\De a_\mu$ to the corresponding component of $\maybetildecgamma$,
the latter being the effective coupling
which describes the decay $\maybepseudoscalar\to\gamma\gamma$.
Each ratio could be expressed as a simple function of the masses appearing in the loops
thanks to the cancellation of the coupling constants common in the numerator and denominator.

We found:
(1) if the 750 GeV resonance does not directly couple to the muon,
the deviation $\De a_\mu$ is difficult to explain;
(2) if it has an enhanced coupling with the muon for instance
as in the 2HDM Type-II or -X, then $\De a_\mu$ can be easily accommodated.
These observations bring us to the following enthralling interpretation of
the latest LHC data:
in conjunction with $\De a_\mu$, the 750 GeV diphoton excess
suggests that the Standard Model Higgs sector should be extended by
\emph{more than} $SU(2)_L$ singlet scalar fields.

\vspace{1ex}
We thank Dominik Stöckinger
for the insightful comments on related questions.
This work is supported in part
by National Research Foundation of Korea (NRF) Research Grant NRF-2015R1A2A1A05001869 (SB).


\begin{thebibliography}{99}
\bibitem{ATLAS:2015}
  The ATLAS collaboration,
  ATLAS-CONF-2015-081.

\bibitem{CMS:2015}
  The CMS Collaboration,
  CMS-PAS-EXO-15-004.

\bibitem{Franceschini:2015kwy}
  R.~Franceschini {\it et al.},
  arXiv:1512.04933 [hep-ph].

\bibitem{Bauer:2015boy}
  M.~Bauer and M.~Neubert,
  arXiv:1512.06828 [hep-ph].

\bibitem{Badziak:2015zez}
  M.~Badziak,
  arXiv:1512.07497 [hep-ph].

\bibitem{Goertz:2015nkp}
  F.~Goertz, J.~F.~Kamenik, A.~Katz and M.~Nardecchia,
  arXiv:1512.08500 [hep-ph].

\bibitem{Nomura:2016fzs}
  T.~Nomura and H.~Okada,
  doi:10.1016/j.physletb.2016.02.022
  arXiv:1601.00386 [hep-ph].

\bibitem{Djouadi:2016eyy}
  A.~Djouadi, J.~Ellis, R.~Godbole and J.~Quevillon,
  arXiv:1601.03696 [hep-ph].

\bibitem{Han:2016bvl}
  X.~F.~Han, L.~Wang and J.~M.~Yang,
  arXiv:1601.04954 [hep-ph].


\bibitem{DiChiara:2015vdm}
  S.~Di Chiara, L.~Marzola and M.~Raidal,
  arXiv:1512.04939 [hep-ph].
%
\bibitem{Altmannshofer:2015xfo}
  W.~Altmannshofer, J.~Galloway, S.~Gori, A.~L.~Kagan, A.~Martin and J.~Zupan,
  arXiv:1512.07616 [hep-ph].


\bibitem{singlet}
  M.~Backovic, A.~Mariotti and D.~Redigolo,
  arXiv:1512.04917 [hep-ph];
%
  D.~Buttazzo, A.~Greljo and D.~Marzocca,
  arXiv:1512.04929 [hep-ph];
%
  J.~Ellis, S.~A.~R.~Ellis, J.~Quevillon, V.~Sanz and T.~You,
  arXiv:1512.05327 [hep-ph];
%
  S.~D.~McDermott, P.~Meade and H.~Ramani,
  arXiv:1512.05326 [hep-ph];
%
  Q.~H.~Cao, Y.~Liu, K.~P.~Xie, B.~Yan and D.~M.~Zhang,
  arXiv:1512.05542 [hep-ph];
%
  A.~Kobakhidze, F.~Wang, L.~Wu, J.~M.~Yang and M.~Zhang,
  arXiv:1512.05585 [hep-ph];
%
  R.~Martinez, F.~Ochoa and C.~F.~Sierra,
  arXiv:1512.05617 [hep-ph];
%
  W.~Chao, R.~Huo and J.~H.~Yu,
  arXiv:1512.05738 [hep-ph];
%
  D.~Curtin and C.~B.~Verhaaren,
  arXiv:1512.05753 [hep-ph];
%
  L.~Bian, N.~Chen, D.~Liu and J.~Shu,
  arXiv:1512.05759 [hep-ph];
%
  A.~Falkowski, O.~Slone and T.~Volansky,
  arXiv:1512.05777 [hep-ph];
%
  R.~Benbrik, C.~H.~Chen and T.~Nomura,
  arXiv:1512.06028 [hep-ph];
%
  H.~Han, S.~Wang and S.~Zheng,
  arXiv:1512.06562 [hep-ph];
%
  W.~Liao and H.~q.~Zheng,
  arXiv:1512.06741 [hep-ph];
%
  D.~Barducci, A.~Goudelis, S.~Kulkarni and D.~Sengupta,
  arXiv:1512.06842 [hep-ph];
%
  I.~Chakraborty and A.~Kundu,
  arXiv:1512.06508 [hep-ph];
%
  O.~Antipin, M.~Mojaza and F.~Sannino,
  arXiv:1512.06708 [hep-ph];
%
  J.~Cao, C.~Han, L.~Shang, W.~Su, J.~M.~Yang and Y.~Zhang,
  arXiv:1512.06728 [hep-ph];
%
  C.~W.~Murphy,
  arXiv:1512.06976 [hep-ph];
%
  K.~Cheung, P.~Ko, J.~S.~Lee, J.~Park and P.~Y.~Tseng,
  arXiv:1512.07853 [hep-ph];
%
  J.~Zhang and S.~Zhou,
  arXiv:1512.07889 [hep-ph];
%
  H.~Han, S.~Wang and S.~Zheng,
  arXiv:1512.07992 [hep-ph];
%
  G.~Li, Y.~n.~Mao, Y.~L.~Tang, C.~Zhang, Y.~Zhou and S.~h.~Zhu,
  arXiv:1512.08255 [hep-ph];
%
  H.~An, C.~Cheung and Y.~Zhang,
  arXiv:1512.08378 [hep-ph];
%
  W.~Chao,
  arXiv:1512.08484 [hep-ph];
%
  S.~Kanemura, N.~Machida, S.~Odori and T.~Shindou,
  arXiv:1512.09053 [hep-ph];
%
  I.~Low and J.~Lykken,
  arXiv:1512.09089 [hep-ph];
%
  A.~E.~C.~Hernández,
  arXiv:1512.09092 [hep-ph];
%
  S.~Jung, J.~Song and Y.~W.~Yoon,
  arXiv:1601.00006 [hep-ph];
%
  T.~Modak, S.~Sadhukhan and R.~Srivastava,
  arXiv:1601.00836 [hep-ph];
%
  F.~D'Eramo, J.~de Vries and P.~Panci,
  arXiv:1601.01571 [hep-ph];
%
  S.~Bhattacharya, S.~Patra, N.~Sahoo and N.~Sahu,
  arXiv:1601.01569 [hep-ph];
%
  P.~Ko and T.~Nomura,
  arXiv:1601.02490 [hep-ph];
%
  W.~Chao,
  arXiv:1601.04678 [hep-ph];
%
  H.~Okada and K.~Yagyu,
  arXiv:1601.05038 [hep-ph];
%
  S.~F.~Ge, H.~J.~He, J.~Ren and Z.~Z.~Xianyu,
  arXiv:1602.01801 [hep-ph];
%
  K.~J.~Bae, M.~Endo, K.~Hamaguchi and T.~Moroi,
  arXiv:1602.03653 [hep-ph].
%

\bibitem{2HDM VLF}
  A.~Angelescu, A.~Djouadi and G.~Moreau,
  arXiv:1512.04921 [hep-ph];
%
  R.~S.~Gupta, S.~Jäger, Y.~Kats, G.~Perez and E.~Stamou,
  arXiv:1512.05332 [hep-ph];
%
  N.~Bizot, S.~Davidson, M.~Frigerio and J.-L.~Kneur,
  arXiv:1512.08508 [hep-ph];
%
  S.~K.~Kang and J.~Song,
  arXiv:1512.08963 [hep-ph].


\bibitem{2HDM}
  D.~Becirevic, E.~Bertuzzo, O.~Sumensari and R.~Z.~Funchal,
  arXiv:1512.05623 [hep-ph];
%
  X.~F.~Han and L.~Wang,
  arXiv:1512.06587 [hep-ph];
%
  W.~C.~Huang, Y.~L.~S.~Tsai and T.~C.~Yuan,
  arXiv:1512.07268 [hep-ph];
%
  X.~F.~Han, L.~Wang, L.~Wu, J.~M.~Yang and M.~Zhang,
  arXiv:1601.00534 [hep-ph];
%
  C.~Arbeláez, A.~E.~C.~Hernández, S.~Kovalenko and I.~Schmidt,
  arXiv:1602.03607 [hep-ph].


\bibitem{others diphoton}
  Y.~Mambrini, G.~Arcadi and A.~Djouadi,
  arXiv:1512.04913 [hep-ph];
%
  K.~Harigaya and Y.~Nomura,
  Phys.\ Lett.\ B {\bf 754} (2016) 151
  doi:10.1016/j.physletb.2016.01.026
  [arXiv:1512.04850 [hep-ph]];
%
  Y.~Nakai, R.~Sato and K.~Tobioka,
  arXiv:1512.04924 [hep-ph];
%
  S.~Knapen, T.~Melia, M.~Papucci and K.~Zurek,
  arXiv:1512.04928 [hep-ph];
%
  A.~Pilaftsis,
  Phys.\ Rev.\ D {\bf 93} (2016) 1,  015017
  doi:10.1103/PhysRevD.93.015017
  [arXiv:1512.04931 [hep-ph]];
%
  B.~Bellazzini, R.~Franceschini, F.~Sala and J.~Serra,
  arXiv:1512.05330 [hep-ph];
%
  E.~Molinaro, F.~Sannino and N.~Vignaroli,
  arXiv:1512.05334 [hep-ph];
%
  T.~Higaki, K.~S.~Jeong, N.~Kitajima and F.~Takahashi,
  Phys.\ Lett.\ B {\bf 755} (2016) 13
  doi:10.1016/j.physletb.2016.01.055
  [arXiv:1512.05295 [hep-ph]];
%
  M.~Low, A.~Tesi and L.~T.~Wang,
  arXiv:1512.05328 [hep-ph];
%
  C.~Petersson and R.~Torre,
  arXiv:1512.05333 [hep-ph];
%
  B.~Dutta, Y.~Gao, T.~Ghosh, I.~Gogoladze and T.~Li,
  arXiv:1512.05439 [hep-ph];
%
  S.~Matsuzaki and K.~Yamawaki,
  arXiv:1512.05564 [hep-ph];
%
  P.~Cox, A.~D.~Medina, T.~S.~Ray and A.~Spray,
  arXiv:1512.05618 [hep-ph];
%
  A.~Ahmed, B.~M.~Dillon, B.~Grzadkowski, J.~F.~Gunion and Y.~Jiang,
  arXiv:1512.05771 [hep-ph];
%
  P.~Agrawal, J.~Fan, B.~Heidenreich, M.~Reece and M.~Strassler,
  arXiv:1512.05775 [hep-ph];
%
  J.~M.~No, V.~Sanz and J.~Setford,
  arXiv:1512.05700 [hep-ph];
%
  S.~V.~Demidov and D.~S.~Gorbunov,
  arXiv:1512.05723 [hep-ph];
%
  S.~Fichet, G.~von Gersdorff and C.~Royon,
  arXiv:1512.05751 [hep-ph];
%
  J.~Chakrabortty, A.~Choudhury, P.~Ghosh, S.~Mondal and T.~Srivastava,
  arXiv:1512.05767 [hep-ph];
%
  C.~Csáki, J.~Hubisz and J.~Terning,
  Phys.\ Rev.\ D {\bf 93} (2016) 3,  035002
  doi:10.1103/PhysRevD.93.035002
  [arXiv:1512.05776 [hep-ph]];
%
  D.~Aloni, K.~Blum, A.~Dery, A.~Efrati and Y.~Nir,
  arXiv:1512.05778 [hep-ph];
%
  Y.~Bai, J.~Berger and R.~Lu,
  arXiv:1512.05779 [hep-ph];
%
  J.~S.~Kim, J.~Reuter, K.~Rolbiecki and R.~R.~de Austri,
  arXiv:1512.06083 [hep-ph];
%
  E.~Gabrielli, K.~Kannike, B.~Mele, M.~Raidal, C.~Spethmann and H.~Veermäe,
  arXiv:1512.05961 [hep-ph];
%
  A.~Alves, A.~G.~Dias and K.~Sinha,
  arXiv:1512.06091 [hep-ph];
%
  E.~Megias, O.~Pujolas and M.~Quiros,
  arXiv:1512.06106 [hep-ph];
%
  L.~M.~Carpenter, R.~Colburn and J.~Goodman,
  arXiv:1512.06107 [hep-ph];
%
  J.~Bernon and C.~Smith,
  arXiv:1512.06113 [hep-ph];
%
  W.~Chao,
  arXiv:1512.06297 [hep-ph];
%
  M.~T.~Arun and P.~Saha,
  arXiv:1512.06335 [hep-ph];
%
  C.~Han, H.~M.~Lee, M.~Park and V.~Sanz,
  doi:10.1016/j.physletb.2016.02.040
  arXiv:1512.06376 [hep-ph];
%
  S.~Chang,
  arXiv:1512.06426 [hep-ph];
%
  M.~Dhuria and G.~Goswami,
  arXiv:1512.06782 [hep-ph];
%
  M.~x.~Luo, K.~Wang, T.~Xu, L.~Zhang and G.~Zhu,
  arXiv:1512.06670 [hep-ph];
%
  J.~Chang, K.~Cheung and C.~T.~Lu,
  arXiv:1512.06671 [hep-ph];
%
  D.~Bardhan, D.~Bhatia, A.~Chakraborty, U.~Maitra, S.~Raychaudhuri and T.~Samui,
  arXiv:1512.06674 [hep-ph];
%
  T.~F.~Feng, X.~Q.~Li, H.~B.~Zhang and S.~M.~Zhao,
  arXiv:1512.06696 [hep-ph];
%
  W.~S.~Cho, D.~Kim, K.~Kong, S.~H.~Lim, K.~T.~Matchev, J.~C.~Park and M.~Park,
  arXiv:1512.06824 [hep-ph];
%
  R.~Ding, L.~Huang, T.~Li and B.~Zhu,
  arXiv:1512.06560 [hep-ph];
%
  F.~Wang, L.~Wu, J.~M.~Yang and M.~Zhang,
  arXiv:1512.06715 [hep-ph];
%
  F.~P.~Huang, C.~S.~Li, Z.~L.~Liu and Y.~Wang,
  arXiv:1512.06732 [hep-ph];
%
  J.~J.~Heckman,
  arXiv:1512.06773 [hep-ph];
%
  X.~J.~Bi, Q.~F.~Xiang, P.~F.~Yin and Z.~H.~Yu,
  arXiv:1512.06787 [hep-ph];
%
  J.~S.~Kim, K.~Rolbiecki and R.~R.~de Austri,
  arXiv:1512.06797 [hep-ph];
%
  L.~Berthier, J.~M.~Cline, W.~Shepherd and M.~Trott,
  arXiv:1512.06799 [hep-ph];
%
  J.~M.~Cline and Z.~Liu,
  arXiv:1512.06827 [hep-ph];
%
  M.~Chala, M.~Duerr, F.~Kahlhoefer and K.~Schmidt-Hoberg,
  Phys.\ Lett.\ B {\bf 755} (2016) 145
  doi:10.1016/j.physletb.2016.02.006
  [arXiv:1512.06833 [hep-ph]];
%
  K.~Kulkarni,
  arXiv:1512.06836 [hep-ph];
%
  S.~M.~Boucenna, S.~Morisi and A.~Vicente,
  arXiv:1512.06878 [hep-ph];
%
  P.~S.~B.~Dev and D.~Teresi,
  arXiv:1512.07243 [hep-ph];
%
  J.~de Blas, J.~Santiago and R.~Vega-Morales,
  arXiv:1512.07229 [hep-ph];
%
  A.~E.~C.~Hernández and I.~Nisandzic,
  arXiv:1512.07165 [hep-ph];
%
  U.~K.~Dey, S.~Mohanty and G.~Tomar,
  arXiv:1512.07212 [hep-ph];
%
  G.~M.~Pelaggi, A.~Strumia and E.~Vigiani,
  arXiv:1512.07225 [hep-ph];
%
  A.~Belyaev, G.~Cacciapaglia, H.~Cai, T.~Flacke, A.~Parolini and H.~Serôdio,
  arXiv:1512.07242 [hep-ph];
%
  M.~Chabab, M.~Capdequi-Peyranère and L.~Rahili,
  arXiv:1512.07280 [hep-ph];
%
  Q.~H.~Cao, S.~L.~Chen and P.~H.~Gu,
  arXiv:1512.07541 [hep-ph];
%
  J.~Gu and Z.~Liu,
  arXiv:1512.07624 [hep-ph];
%
  S.~Moretti and K.~Yagyu,
  arXiv:1512.07462 [hep-ph];
%
  K.~M.~Patel and P.~Sharma,
  arXiv:1512.07468 [hep-ph];
%
  S.~Chakraborty, A.~Chakraborty and S.~Raychaudhuri,
  arXiv:1512.07527 [hep-ph];
%
  M.~Cvetič, J.~Halverson and P.~Langacker,
  arXiv:1512.07622 [hep-ph];
%
  B.~C.~Allanach, P.~S.~B.~Dev, S.~A.~Renner and K.~Sakurai,
  arXiv:1512.07645 [hep-ph];
%
  H.~Davoudiasl and C.~Zhang,
  arXiv:1512.07672 [hep-ph];
%
  K.~Das and S.~K.~Rai,
  arXiv:1512.07789 [hep-ph];
%
  N.~Craig, P.~Draper, C.~Kilic and S.~Thomas,
  arXiv:1512.07733 [hep-ph];
%
  J.~Liu, X.~P.~Wang and W.~Xue,
  arXiv:1512.07885 [hep-ph];
%
  J.~A.~Casas, J.~R.~Espinosa and J.~M.~Moreno,
  arXiv:1512.07895 [hep-ph];
%
  L.~J.~Hall, K.~Harigaya and Y.~Nomura,
  arXiv:1512.07904 [hep-ph];
%
  J.~C.~Park and S.~C.~Park,
  arXiv:1512.08117 [hep-ph];
%
  D.~Chway, R.~Dermíšek, T.~H.~Jung and H.~D.~Kim,
  arXiv:1512.08221 [hep-ph];
%
  A.~Salvio and A.~Mazumdar,
  arXiv:1512.08184 [hep-ph];
%
  M.~Son and A.~Urbano,
  arXiv:1512.08307 [hep-ph];
%
  F.~Wang, W.~Wang, L.~Wu, J.~M.~Yang and M.~Zhang,
  arXiv:1512.08434 [hep-ph];
%
  Q.~H.~Cao, Y.~Liu, K.~P.~Xie, B.~Yan and D.~M.~Zhang,
  arXiv:1512.08441 [hep-ph];
%
  J.~Gao, H.~Zhang and H.~X.~Zhu,
  arXiv:1512.08478 [hep-ph];
%
  P.~S.~B.~Dev, R.~N.~Mohapatra and Y.~Zhang,
  arXiv:1512.08507 [hep-ph];
%
  Y.~L.~Tang and S.~h.~Zhu,
  arXiv:1512.08323 [hep-ph];
%
  J.~Cao, F.~Wang and Y.~Zhang,
  arXiv:1512.08392 [hep-ph];
%
  C.~Cai, Z.~H.~Yu and H.~H.~Zhang,
  arXiv:1512.08440 [hep-ph];
%
  J.~E.~Kim,
  Phys.\ Lett.\ B {\bf 755} (2016) 190
  doi:10.1016/j.physletb.2016.02.016
  [arXiv:1512.08467 [hep-ph]];
%
  X.~J.~Bi {\it et al.},
  arXiv:1512.08497 [hep-ph];
%
  L.~A.~Anchordoqui, I.~Antoniadis, H.~Goldberg, X.~Huang, D.~Lust and T.~R.~Taylor,
  doi:10.1016/j.physletb.2016.02.024
  arXiv:1512.08502 [hep-ph];
%
  L.~E.~Ibanez and V.~Martin-Lozano,
  arXiv:1512.08777 [hep-ph];
%
  Y.~Hamada, T.~Noumi, S.~Sun and G.~Shiu,
  arXiv:1512.08984 [hep-ph];
%
  X.~J.~Huang, W.~H.~Zhang and Y.~F.~Zhou,
  arXiv:1512.08992 [hep-ph];
%
  C.~W.~Chiang, M.~Ibe and T.~T.~Yanagida,
  arXiv:1512.08895 [hep-ph];
%
  S.~Kanemura, K.~Nishiwaki, H.~Okada, Y.~Orikasa, S.~C.~Park and R.~Watanabe,
  arXiv:1512.09048 [hep-ph];
%
  P.~V.~Dong and N.~T.~K.~Ngan,
  arXiv:1512.09073 [hep-ph];
%
  Y.~Jiang, Y.~Y.~Li and T.~Liu,
  arXiv:1512.09127 [hep-ph];
%
  K.~Kaneta, S.~Kang and H.~S.~Lee,
  arXiv:1512.09129 [hep-ph];
%
  L.~Marzola, A.~Racioppi, M.~Raidal, F.~R.~Urban and H.~Veermäe,
  arXiv:1512.09136 [hep-ph];
%
  E.~Ma,
  arXiv:1512.09159 [hep-ph];
%
  A.~Dasgupta, M.~Mitra and D.~Borah,
  arXiv:1512.09202 [hep-ph];
%
  C.~T.~Potter,
  arXiv:1601.00240 [hep-ph];
%
  E.~Palti,
  arXiv:1601.00285 [hep-ph];
%
  P.~Ko, Y.~Omura and C.~Yu,
  arXiv:1601.00586 [hep-ph];
%
  K.~Ghorbani and H.~Ghorbani,
  arXiv:1601.00602 [hep-ph];
%
  U.~Danielsson, R.~Enberg, G.~Ingelman and T.~Mandal,
  arXiv:1601.00624 [hep-ph];
%
  W.~Chao,
  arXiv:1601.00633 [hep-ph];
%
  C.~Csaki, J.~Hubisz, S.~Lombardo and J.~Terning,
  arXiv:1601.00638 [hep-ph];
%
  A.~Karozas, S.~F.~King, G.~K.~Leontaris and A.~K.~Meadowcroft,
  arXiv:1601.00640 [hep-ph];
%
  A.~E.~C.~Hernández, I.~d.~M.~Varzielas and E.~Schumacher,
  arXiv:1601.00661 [hep-ph];
%
  B.~Dutta, Y.~Gao, T.~Ghosh, I.~Gogoladze, T.~Li, Q.~Shafi and J.~W.~Walker,
  arXiv:1601.00866 [hep-ph];
%
  F.~F.~Deppisch, C.~Hati, S.~Patra, P.~Pritimita and U.~Sarkar,
  arXiv:1601.00952 [hep-ph];
%
  H.~Ito, T.~Moroi and Y.~Takaesu,
  arXiv:1601.01144 [hep-ph];
%
  H.~Zhang,
  arXiv:1601.01355 [hep-ph];
%
  A.~Berlin,
  arXiv:1601.01381 [hep-ph];
%
  E.~Ma,
  arXiv:1601.01400 [hep-ph];
%
  I.~Sahin,
  arXiv:1601.01676 [hep-ph];
%
  S.~Fichet, G.~von Gersdorff and C.~Royon,
  arXiv:1601.01712 [hep-ph];
%
  D.~Borah, S.~Patra and S.~Sahoo,
  arXiv:1601.01828 [hep-ph];
%
  D.~Stolarski and R.~Vega-Morales,
  arXiv:1601.02004 [hep-ph];
%
  J.~Cao, L.~Shang, W.~Su, Y.~Zhang and J.~Zhu,
  arXiv:1601.02570 [hep-ph];
%
  M.~Fabbrichesi and A.~Urbano,
  arXiv:1601.02447 [hep-ph];
%
  C.~Hati,
  arXiv:1601.02457 [hep-ph];
%
  J.~H.~Yu,
  arXiv:1601.02609 [hep-ph];
%
  R.~Ding, Z.~L.~Han, Y.~Liao and X.~D.~Ma,
  arXiv:1601.02714 [hep-ph];
%
  S.~Alexander and L.~Smolin,
  arXiv:1601.03091 [hep-ph];
%
  J.~H.~Davis, M.~Fairbairn, J.~Heal and P.~Tunney,
  arXiv:1601.03153 [hep-ph];
%
  L.~V.~Laperashvili, H.~B.~Nielsen and C.~R.~Das,
  arXiv:1601.03231 [hep-ph];
%
  I.~Dorsner, S.~Fajfer and N.~Kosnik,
  arXiv:1601.03267 [hep-ph];
%
  A.~E.~Faraggi and J.~Rizos,
  arXiv:1601.03604 [hep-ph];
%
%
  A.~Ghoshal,
  arXiv:1601.04291 [hep-ph];
%
  T.~Nomura and H.~Okada,
  arXiv:1601.04516 [hep-ph];
%
  M.~R.~Buckley,
  arXiv:1601.04751 [hep-ph];
%
  D.~B.~Franzosi and M.~T.~Frandsen,
  arXiv:1601.05357 [hep-ph];
%
  A.~Martini, K.~Mawatari and D.~Sengupta,
  arXiv:1601.05729 [hep-ph];
%
  Q.~H.~Cao, Y.~Q.~Gong, X.~Wang, B.~Yan and L.~L.~Yang,
  arXiv:1601.06374 [hep-ph];
%
  C.~W.~Chiang and A.~L.~Kuo,
  arXiv:1601.06394 [hep-ph];
%
  U.~Aydemir and T.~Mandal,
  arXiv:1601.06761 [hep-ph];
%
  S.~Abel and V.~V.~Khoze,
  arXiv:1601.07167 [hep-ph];
%
  L.~A.~Harland-Lang, V.~A.~Khoze and M.~G.~Ryskin,
  arXiv:1601.07187 [hep-ph];
%
  S.~F.~King and R.~Nevzorov,
  arXiv:1601.07242 [hep-ph];
%
  J.~Kawamura and Y.~Omura,
  arXiv:1601.07396 [hep-ph];
%
  B.~J.~Kavanagh,
  arXiv:1601.07330 [hep-ph];
%
  T.~Nomura and H.~Okada,
  arXiv:1601.07339 [hep-ph];
%
  C.~Q.~Geng and D.~Huang,
  arXiv:1601.07385 [hep-ph];
%
  E.~Bertuzzo, P.~A.~N.~Machado and M.~Taoso,
  arXiv:1601.07508 [hep-ph];
%
  I.~Ben-Dayan and R.~Brustein,
  arXiv:1601.07564 [hep-ph];
%
  A.~D.~Martin and M.~G.~Ryskin,
  J.\ Phys.\ G {\bf 43} (2016) 04
  doi:10.1088/0954-3899/43/4/04LT02
  [arXiv:1601.07774 [hep-ph]];
%
  A.~Hektor and L.~Marzola,
  arXiv:1602.00004 [hep-ph];
%
  N.~D.~Barrie, A.~Kobakhidze, M.~Talia and L.~Wu,
  doi:10.1016/j.physletb.2016.02.010
  arXiv:1602.00475 [hep-ph];
%
  L.~Aparicio, A.~Azatov, E.~Hardy and A.~Romanino,
  arXiv:1602.00949 [hep-ph];
%
  R.~Ding, Y.~Fan, L.~Huang, C.~Li, T.~Li, S.~Raza and B.~Zhu,
  arXiv:1602.00977 [hep-ph];
%
  K.~Harigaya and Y.~Nomura,
  arXiv:1602.01092 [hep-ph];
%
  T.~Li, J.~A.~Maxin, V.~E.~Mayes and D.~V.~Nanopoulos,
  arXiv:1602.01377 [hep-ph];
%
  A.~Salvio, F.~Staub, A.~Strumia and A.~Urbano,
  arXiv:1602.01460 [hep-ph];
%
  S.~I.~Godunov, A.~N.~Rozanov, M.~I.~Vysotsky and E.~V.~Zhemchugov,
  arXiv:1602.02380 [hep-ph];
%
%
  S.~B.~Giddings and H.~Zhang,
  arXiv:1602.02793 [hep-ph];
%
  U.~Ellwanger and C.~Hugonie,
  arXiv:1602.03344 [hep-ph];
%
  P.~Draper and D.~McKeen,
  arXiv:1602.03604 [hep-ph];
%
  C.~Gross, O.~Lebedev and J.~M.~No,
  arXiv:1602.03877 [hep-ph];
%
  C.~Han, T.~T.~Yanagida and N.~Yokozaki,
  arXiv:1602.04204 [hep-ph];
%
  Y.~Hamada, H.~Kawai, K.~Kawana and K.~Tsumura,
  arXiv:1602.04170 [hep-ph];
%
  F.~Goertz, A.~Katz, M.~Son and A.~Urbano,
  arXiv:1602.04801 [hep-ph];
%
  B.~Dasgupta, J.~Kopp and P.~Schwaller,
  arXiv:1602.04692 [hep-ph];
%
  C.~Frugiuele, E.~Fuchs, G.~Perez and M.~Schlaffer,
  arXiv:1602.04822 [hep-ph];
%
  C.~Delaunay and Y.~Soreq,
  arXiv:1602.04838 [hep-ph];
%
  Y.~J.~Zhang, B.~B.~Zhou and J.~J.~Sun,
  arXiv:1602.05539 [hep-ph].

\bibitem{g-2}
 F.~Jegerlehner and A.~Nyffeler,
  Phys.\ Rept.\  {\bf 477} (2009) 1
  doi:10.1016/j.physrep.2009.04.003
  [arXiv:0902.3360 [hep-ph]];
M.~Benayoun, P.~David, L.~DelBuono and F.~Jegerlehner,
  Eur.\ Phys.\ J.\ C {\bf 73} (2013) 2453
  doi:10.1140/epjc/s10052-013-2453-3
  [arXiv:1210.7184 [hep-ph]].

\bibitem{Barr:1990vd}
  S.~M.~Barr and A.~Zee,
  Phys.\ Rev.\ Lett.\  {\bf 65} (1990) 21
   Erratum: [Phys.\ Rev.\ Lett.\  {\bf 65} (1990) 2920].
  doi:10.1103/PhysRevLett.65.21

\bibitem{Arhrib:2001}
  A.~Arhrib and S.~Baek,
  Phys.\ Rev.\ D {\bf 65} (2002) 075002
  doi:10.1103/PhysRevD.65.075002
  [hep-ph/0104225].

\bibitem{Baek:2002}
  S.~Baek, P.~Ko and J.-h.~Park,
  Eur.\ Phys.\ J.\ C {\bf 24} (2002) 613
  doi:10.1007/s10052-002-0971-5
  [hep-ph/0203251].

\bibitem{Baek:2004tm}
  S.~Baek,
  Phys.\ Lett.\ B {\bf 595} (2004) 461
  doi:10.1016/j.physletb.2004.05.071
  [hep-ph/0406007].

\bibitem{Baek:2014}
  S.~Baek,
  JHEP {\bf 1508} (2015) 023
  doi:10.1007/JHEP08(2015)023
  [arXiv:1410.1992 [hep-ph]].

\bibitem{Ilisie:2015}
  V.~Ilisie,
  JHEP {\bf 1504} (2015) 077
  doi:10.1007/JHEP04(2015)077
  [arXiv:1502.04199 [hep-ph]].

\bibitem{Abe:2015oca}
  T.~Abe, R.~Sato and K.~Yagyu,
  JHEP {\bf 1507} (2015) 064
  doi:10.1007/JHEP07(2015)064
  [arXiv:1504.07059 [hep-ph]].

\bibitem{Falkowski:2012}
  D.~Carmi, A.~Falkowski, E.~Kuflik, T.~Volansky and J.~Zupan,
  JHEP {\bf 1210} (2012) 196
  doi:10.1007/JHEP10(2012)196
  [arXiv:1207.1718 [hep-ph]].

\bibitem{Djouadi:1993ji}
  A.~Djouadi, M.~Spira and P.~M.~Zerwas,
  Phys.\ Lett.\ B {\bf 311} (1993) 255
  doi:10.1016/0370-2693(93)90564-X
  [hep-ph/9305335].

\bibitem{LHCHiggs}
  The ATLAS and CMS Collaborations,
  ATLAS-CONF-2015-044.

\bibitem{Chpoi:2013wga}
S.~Baek, P.~Ko and W.~I.~Park,
  JHEP {\bf 1202} (2012) 047
  doi:10.1007/JHEP02(2012)047
  [arXiv:1112.1847 [hep-ph]];
  S.~Choi, S.~Jung and P.~Ko,
  JHEP {\bf 1310} (2013) 225
  doi:10.1007/JHEP10(2013)225
  [arXiv:1307.3948 [hep-ph]].

\bibitem{Weinberg:1976hu}
  S.~Weinberg,
  Phys.\ Rev.\ Lett.\  {\bf 37} (1976) 657.
  doi:10.1103/PhysRevLett.37.657

\bibitem{Glashow:1976nt}
  S.~L.~Glashow and S.~Weinberg,
  Phys.\ Rev.\ D {\bf 15} (1977) 1958.
  doi:10.1103/PhysRevD.15.1958

\bibitem{Aoki:2009ha}
  M.~Aoki, S.~Kanemura, K.~Tsumura and K.~Yagyu,
  Phys.\ Rev.\ D {\bf 80} (2009) 015017
  doi:10.1103/PhysRevD.80.015017
  [arXiv:0902.4665 [hep-ph]].

\bibitem{Craig:2012vn}
  N.~Craig and S.~Thomas,
  JHEP {\bf 1211} (2012) 083
  doi:10.1007/JHEP11(2012)083
  [arXiv:1207.4835 [hep-ph]].

\bibitem{Aad:2014vgg}
  G.~Aad {\it et al.} [ATLAS Collaboration],
  JHEP {\bf 1411} (2014) 056
  doi:10.1007/JHEP11(2014)056
  [arXiv:1409.6064 [hep-ex]].

\bibitem{Misiak:2015xwa}
  M.~Misiak {\it et al.},
  Phys.\ Rev.\ Lett.\  {\bf 114} (2015) no.22,  221801
  doi:10.1103/PhysRevLett.114.221801
  [arXiv:1503.01789 [hep-ph]].

\bibitem{Hermann:2012fc}
  T.~Hermann, M.~Misiak and M.~Steinhauser,
  JHEP {\bf 1211} (2012) 036
  doi:10.1007/JHEP11(2012)036
  [arXiv:1208.2788 [hep-ph]].

\bibitem{Chun:2015xfx}
  E.~J.~Chun,
  arXiv:1511.05225 [hep-ph].

\end{thebibliography}
\end{document}